\documentclass[12pt,useAMS,usenatbib]{article} \usepackage{graphics}
\usepackage{amsmath} \usepackage{jcappub}
\usepackage{natbib} \usepackage{hyperref} \usepackage{color}
\usepackage{psfrag} \usepackage{epsfig}

\title{\boldmath Predictions for measuring the cross power spectrum of
  the HI 21-cm signal and the Lyman-$\alpha$ forest using OWFA}

\author[a,1]{Anjan Kumar Sarkar, \note{Corresponding author.}}
\author[a,b,2]{Somnath Bharadwaj,} 
\author[c,3]{Tapomoy Guha Sarkar}

\affiliation[a]{Centre for Theoretical Studies, IIT Kharagpur, Kharagpur - 721302, India} 
\affiliation[b]{Department of Physics, IIT Kharagpur, Kharagpur - 721302, India} 
\affiliation[c]{Birla Institute of Technology and Science, Pilani - 333031, India}

\emailAdd{anjan@cts.iitkgp.ernet.in}
\emailAdd{somnath@phy.iitkgp.ernet.in}
\emailAdd{tapomoy@pilani.bits-pilani.ac.in}

\abstract{We have studied the possibility of measuring the
  cross-correlation of the redshifted HI 21-cm signal and the
  Lyman-$\alpha$ forest using an upcoming radio-interferometric array
  OWFA and an spectroscopic observation like SDSS-IV. Our results
  shows that it is possible to have a $6 \sigma$ detection of the
  cross-correlation signal with OWFA PII using an observing time of
  $200$ hrs each in $N_p = 25$ independent fields-of-view. However,
  not much can be done beyond this using the cross-correlation signal
  for $z_c = 3.35$ and $B = 30 \, {\rm MHz}$. Apart from this, we have
  also envisaged a situation where observations are carried out at
  $z_c = 3.05$ and $2.55$ which lie closer to the peak of the quasar
  distribution at $z = 2.25$ and with a larger bandwidth of $B = 60 \,
  {\rm MHz}$. We see that the SNR of the cross-correlation detection
  can be significantly enhanced to $\sim 17$ for $z_c = 2.55$ and $B =
  60 \, {\rm MHz}$. It is then possible to measure $\beta_T$ and
  $\beta_F$ individually with an ${\rm SNR} \ge 5$ by combining the
  cross-correlation with the HI 21-cm auto-correlation
  measurements. We further find that a measurement of the binned
  cross-correlation power spectrum with ${\rm SNR} \ge 5$ is also
  possible in several bins at $k \le 0.3 \, {\rm Mpc}^{-1}$.}

\begin{document}
\maketitle

\section{Introduction}
 
Observation of the redshifted 21-cm signal from Neutral Hydrogen (HI) is considered to be a promising tool to map out the large scale structure of the universe from the post-reionization era ($z \leq 6$). Here, majority of the HI is hosted by the discrete sources which have the column number density, $N_{\rm HI} \geq 2 \times 10^{20}$ atoms/cm$^{2}$ \citep{lanzetta-wolfe1995,zafar2013}. Emission from these sources appears as the diffuse background in low frequency radio observations below 1420 MHz. Fluctuations in the background radiation in angular and the frequency scales, which are quantified through the HI power spectrum, carries the signature of the underlying source clustering at that epoch and can thereby be used to probe the large scale structure of the universe at high $z$ \citep{BNS2001,bharadwaj-sethi2001,wyithe-loeb2008}. Measurement of the HI power spectrum holds the potential of measuring the Baryon Acoustic Oscillations (BAO), which can be used to constrain the models of the dark energy \citep{Wyithe-geil2008,chang2008,seo2010,ansari-bao2012}. The measurement of the HI power spectrum can be used further to constrain the background cosmological model \citep{bharadwaj2009,visbal2009} and the neutrino mass \citep{wyithe-loeb2008,villaescusa2015,pal2016}. In addition to the HI power spectrum, these fluctuations can also be used to probe the non-Gaussinities in the HI 21-cm signal, such as bispectrum \citep{ali-bispectrum2006,sarkar-bispectrum2013}.

The Lyman-$\alpha$ forest are identified as the series of multiple
Lyman-$\alpha$ absorption lines seen in the spectra of the backgrounds
quasars. These absorption features arise from the tiny fluctuations in the HI
density of the predominantly ionized diffuse IGM. Observations of the
Lyman-$\alpha$ forest trace out the HI distribution in the direction of the
line-of-sight of the observed quasars. On suitably large scales, the fluctuations in the HI 21-cm signal
and transmitted flux of the quasars through the Lyman-$\alpha$ forest are
expected to be a biased tracer of the matter distribution on large
scales. Alike the Hi 21-cm signal, the observations of the Lyman-$\alpha$
forest also finds use in a host of areas, such as the measurement of the
matter power spectrum \citep{croft-weinberg1998lyman,croft-weinberg1999lyman,mcdonald2005lyman} and bispectrum \citep{mandelbaum2003lyman,croft-weinberg2002lyman,vielbispectrum2004lyman}, cosmological parameters \citep{bi-davidson1997lyman,kim2004lyman}, constraints on neutrino mass \citep{croft-neutrino1999lyman,yeche2017lyman}, dark energy \citep{mcdonald2007lyman} and reionization history \citep{gallerani2006reionization}. In addition to these, the recent Baryon Acoustic Oscillations Survey (BOSS) targets to probe the
dark energy and the cosmic acceleration by measuring the large scale structure
and imprints of BAO in the Lyman-$\alpha$ forest \citep{delubac2013lyman,delubac2015lyman}.

The cross correlation of the HI 21-cm signal and the Lyman-$\alpha$ forest has
been proposed to be a potential probe of the post-reionization era
\citep{guha-sarkar2010}. The cross correlation signal of the redshifted HI
21-cm emission and the Lyman-$\alpha$ forest carries a few unique features
that makes it worthwhile to consider it a viable probe of the HI power
spectrum apart from the auto-correlation of the individual signals. We begin
with a brief discussion on a few aspects of the autocorrelation of the
Lyman-$\alpha$ forest and the HI 21-cm signal to emphasize the features
particular to the cross correlation signal. In the case of  Lyman-$\alpha$
forest observations, the spectra  can only be detected at discrete line of sights to the 
quasars. The Poisson noise due to the discrete sampling of the quasars, limits
the accuracy with which one would be able to estimate the three dimensional
auto power spectrum from a given survey. This limit is fixed by the quasar
number density and the Signal-to-noise ratio for the each of the individual
quasar spectrum. Both of these factors are specific
to the instrument and observational 
strategy and one can not improve these for a given survey.

 Contrary to the Lyman-$\alpha$ observations, redshifted HI 21-cm observations are sensitive to the HI
distribution within the total field-of-view under observation. The accuracy of
the measurement of the three-dimensional HI power spectrum depends on the
array configuration of the instrument and the system noise. One can, in
principle, use  larger observation time to
improve the precision with which one can measure the HI power
spectrum. The redshifted HI 21-cm signal, however, remains buried in
foregrounds from other sources like, galactic synchrotron radiation, point
sources etc. that are several orders of magnitude larger than the HI signal
\citep{ghosh2011}. Separating the foregrounds from the HI signal poses a
serious 
challenge  towards measuring the HI power spectrum.  This is in
contrast to the cross correlation signal, where foregrounds for the redshifted
HI 21-cm signal and the Lyman-$\alpha$ forest are not correlated with each
other. The problem of the foregrounds are thereby much less severe in case of
the cross correlation signal as compared to that for the auto-correlation of
the redshifted HI 21-cm signal, which gives the cross correlation an advantage
over the redshifted 21-cm autocorrelation. In the presence of
  astrophysical foregrounds, 
  one can  be certain about
  the cosmological nature of the HI signal only if it is detected in a
  cross-correlation. In case of Lyman-$\alpha$
autocorrelation, the precision with which it is possible to measure the power
spectrum is limited by the finite sampling of the QSOs. The issue of finite
sampling of the QSOs is expected to be less sensitive for the cross
correlation signal in comparison with the Lyman-$\alpha$ forest
autocorrelation. Combined with the Lyman-$\alpha$ forest survey, a HI 21-cm
observations can be suitably designed to measure the power spectrum with high
accuracy using the cross correlation signal. The less severe issue of
foregrounds subtraction 
compared to the redshifted HI 21-cm observations and prospects for higher SNR
than the Lyman-$\alpha$ forest auto-correlation, serves the primary motivation
towards using the cross correlation of the HI 21-cm signal and the
Lyman-$\alpha$ forest.

There are a number of studies in the recent few years on using the
cross-correlation of the HI 21-cm signal and the Lyman-$\alpha$ forest as the
viable probe of the cosmology from the post reionization era
\citep{guha-sarkar2010}. The cross-correlation of the HI 21-cm signal and the
Lyman-$\alpha$ forest can be used to measure the matter power spectrum in the
post-reionization era \citep{guha-sarkar2011cross,guha-sarkar2016cross}. 
\citet{guha-sarkar2013} have explored the possibility of using the cross correlation signal to make analytical estimates of the accuracy with which one would be able to detect the Baryon Acoustic
Oscillations (BAO). In a subsequent study, \citet{guha-sarkar2015} have looked
for the possibility of detecting the three-dimensional cross correlation of
the redshifted HI 21-cm signal and the Lyman-$\alpha$ forest using a future
telescope like SKA-mid and a future experiment like Baryon Acoustic
Oscillation Survey (BOSS). \citet{carucci2017} have investigated the cross
correlation signal in the non-linear regime using a complete state-of -the-art
hydrodynamic N-body simulations and have found that the linear theory
predictions for the shape and the amplitude of the cross correlation power
spectrum even holds to a rather non-linear scales.

We have used OWFA to study the prospects of detecting the cross-correlation of the redshifted HI 21-cm signal and the Lyman-$\alpha$ forest from the spectroscopic observations like Baryon Oscillations Spectroscopic Survey or  BOSS. OWFA is an upgradation of the Ooty Radio
Telescope (ORT), which is expected to operate as a linear
radio-interferometric array with a frequency of $\nu_0 = 326.5$ MHz, that
corresponds to measuring the HI radiation from the redshift of $z_0 = 3.35$
\citep{prasad2011,marthi2013}. The ORT is 530 m long and 30 m wide parabolic
cylindrical reflector, that is placed along the north-south direction on a
hill of slope $11^{\circ}$ which is same as the latitude of the place
\citep{swarup1971,sarma1975}. It is thereby possible to observe a fixed part
of the sky with a single rotation of the telescope. The feed system of the ORT
consists of 1056 dipoles, which are placed 
end-to-end at $0.48 \, {\rm m}$ apart from each other along the length of the cylinder. 
OWFA is capable to operate in two simultaneous independent interferometric modes - PI and PII
\citep{ali-bharadwaj2014}. The PI and PII have a total of $40$ antennas and $264$ antennas, 
where signals from $24$ dipoles and $4$ dipoles have been combined to form an single 
antenna element respectively. The operating bandwidth for the PII is $39$ MHz.

A number of studies has been carried out in recent few years to explore the
prospects of the OWFA. \citet{ali-bharadwaj2014}, have studied for the first
time, the prospects for detecting the HI 21-cm signal using OWFA. Their study
also contain detailed foreground predictions for OWFA. In a later study,
\citet{fisher2015} have used the Fisher matrix analysis to make estimates of
the accuracy with which it would be possible to measure the amplitude of the
HI power spectrum. In a recent study, \citet{fisherbin2017} have made
predictions based on Fisher matrix analysis for constraining the shape of the
HI power spectrum over the $k$-range probed by the OWFA phase II.

We have used the Fisher matrix analysis to make an estimate of the accuracy
with which it would be possible to measure the cross correlation power
spectrum using the 21-cm observations with OWFA and the spectroscopic
observation like BOSS of SDSS-IV. 
We have carried out our analysis to study
prospects of measuring the cross correlation power spectrum from two different
aspects. On the first hand, We have computed the expected signal-to-Noise
ratios (SNR) for measuring the amplitude of the cross correlation power
spectrum, $A_{FT}$. We assumed that both the HI distribution traces
the underlying matter power spectrum on large scales and the shape of the
matter power spectrum is specified precisely by the standard $\Lambda{\rm
  CDM}$ cosmological model.
In addition to this, we have also made estimates
for detecting $\beta_T$, redshift space distortion parameter for the HI power
spectrum and $\beta_F$, the redshift space distortion parameter for the
Lyman-$\alpha$ forest. 
This is the first full covariance matrix analysis of the 21-cm and Ly
-$\alpha$ crosscorrelation using  the observationally meaningful visibility based approach.

Further, we have explored the possibility of directly measuring
the binned cross correlation power spectrum without any prior information
about its amplitude and shape.  Several
astrophysical processes can alter the shape of the cross correlation power
spectrum without any reference to the change in the matter power
spectrum. Further, changes in the background cosmological model will be
reflected as the changes in the shape of the matter power spectrum via several
effects like redshift space distortion, Alcock-Paczynski (AP) effect etc. For
this purpose, we have divided the entire $k$-range into a number of $k$-bins
and have used the Fisher matrix analysis to estimate the SNR for measuring the
amplitude in each of these $k$-bins. We have used the transfer function of the
baryonic matter power spectrum of \citet{eisenstein-hu1998} with cosmological
parameters taken from \citet{planck2014}. Constraining the shape of the
cross-correlation power spectrum directly gives us a novel way to measure
various parameters of astrophysical and cosmological interest. 

The present paper is outlined as follows. In section 2, we present a
discussion about the cross correlation power spectrum  between of the HI 21-cm signal
and Lyman-$\alpha$ forest. In section 3, we give a brief overview of the
parameters of the OWFA and have calculate the Fisher matrix of the
cross-correlation of the visibilities of the HI 21-cm signal and the
Lyman-$\alpha$ forest for OWFA. In section 4, we discuss the results of the
Fisher matrix analysis for obtaining constraints on the parameters  $A_{FT}$, $\beta_T$ and $\beta_F$. We have also presented
estimates for measuring the binned power spectrum in this section. The
concluding Section 5 summarizes the results of our analysis.

\section{Cross correlation signal of HI 21-cm and Lyman-$\alpha$ forest}

The brightness temperature fluctuations $\delta_T(\hat{n}, z)$ and the
dimensionless fluctuations $\delta_F(\hat{n}, z)$ of transmitted flux
are the quantities of interest for the redshift HI 21-cm signal and
the Lyman $\alpha$ forest respectively.  Here we consider observation
in a region of angular extent $L \times L$ and redshift band $B$ which
respectively correspond to the field of view (FoV) and the frequency
band of the radio telescope used for the 21-cm observations.  In the
flat-sky approximation, we can express the unit vector $\hat{n}$ as
$\hat{n} = \hat{m} + \vec{\theta}$ where $\vec{\theta}$ is a
two-dimensional vector on the plane of the sky and $\hat{m}$ is a unit
vector along the line-of-sight to the centre of the FoV. We further
assume that the band $B$ is divided into $N_c$ number of channels each
of width $\Delta z_c$ (which corresponds to the frequency channel
width of the radio telescope), and we can express $z= z_c + \, \Delta
z_n$ where $z_c$ is the redshift at the center of the band, $\Delta
z_n= n \Delta z_c$ and $-N_c/2 < n \leq +N_c/2$.

It is convenient to decompose the fluctuations
$\delta_{\alpha}(\vec{\theta}, \Delta z_n)$ (here $\alpha = F,T$) into
Fourier modes using \citep{guha-sarkar2013}
\begin{equation}
\delta_{\alpha}(\vec{\theta}, \Delta z_n) = (B L^2)^{-1} \,
\sum_{\tau} \, \sum_{\vec{U}} \, \Delta_{\alpha} (\vec{U}, \tau) \,
e^{2 \pi i (\vec{U} \cdot \vec{\theta} + \tau \Delta z_n)}
\label{eq:a1}
\end{equation}
where $\vec{U}$ and $\tau$ are the Fourier conjugates of
$\vec{\theta}$ $\Delta z_n$ respectively.

We quantify the statistical properties of these fluctuations using the
power spectra $\mathcal{P}_{\alpha \gamma} (\vec{U}, \tau_n)$ which, n
the limit of $L \rightarrow \infty$, are defined through
\begin{equation}
\langle \Delta_{\alpha} (\vec{U}, \tau_m) \, \Delta^{*}_{\gamma}
(\vec{U}^{'}, \tau_n) = B \, \delta_{m,n} \, \delta^2(\vec{U} -
\vec{U}^{'}) \, \mathcal{P}_{\alpha \gamma} (U, \tau_n)
\label{eq:pk}
\end{equation}
where $\delta_{m,n}$ is the Kronecker delta and $\delta^2(\vec{U} -
\vec{U}^{'})$ is the Dirac delta function. Here the power spectrum
$\mathcal{P}_{TT}(U, \tau_m)$ and $\mathcal{P}_{FF}(U, \tau_m)$
respectively refer to the 21-cm and the Lyman-$\alpha$ forest
auto-correlation signals whereas $\mathcal{P}_{TF}(U, \tau_m)$ refers
to cross-correlation between these two signals.  Note that we have
discrete values $\tau_m = m/B$ with $-N_c/2 < m \le N_c/2$.

 On sufficiently large length-scales it is reasonable to assume that
 the 21-cm and the Lyman-$\alpha$ forest signals are both linearly
 related to the underlying matter fluctuations whereby we can express
 the power spectra $\mathcal{P}_{\alpha \gamma}(U, \tau_m)$ as
\begin{equation}
\mathcal{P}_{\alpha \gamma} (U, \tau_m) = F_{\alpha \gamma} (\mu) \,
P(k)
\label{eq:PFT}
\end{equation}
where $P(k)$ is the dark matter power spectrum and the comoving wave
vector $\vec{k}$ with $k = \mid \vec{k} \mid$ has components
$\vec{k}_{\perp}=2 \pi \vec{U}/r$ and $k_{\parallel} = 2 \pi H(z)
\tau_m/c$ respectively perpendicular and parallel to the line of
sight. Here $r$ and $H(z)$ respectively refer to the co-moving
distance and the Hubble parameter corresponding to $z_c$ and
\begin{equation}
F_{\alpha \gamma} (\mu) = H(z)(c r^2)^{-1} \, C_{\alpha} \, C_{\gamma}
\, (1+\beta_{\alpha} \mu^2) \, (1+\beta_{\gamma} \mu^2)
\label{eq:FF}
\end{equation} 
Where $\mu = k_{\parallel}/k$. For the Lyman-$\alpha$ forest we have
used the values $C_F =(-0.13,-0.29,-0.29)$ and $\beta_F = (
1.58,1.13,1.13)$ at redshifts $z_c=(2.55,3.05,3.35)$ respectively, this is based
on the fit to the 1D Lyman-$\alpha$ power spectrum presented in
\citet{Palanque2013}.  For the HI 21-cm signal we have used $C_T =
(\bar{T} x_{HI} b_{HI})$ and $\beta_T=f(\Omega)/b_{HI}$ where
$\bar{T}(z)$ is the characteristic 21-cm brightness temperature
\citep{bharadwaj-ali2005}, $f(\Omega)$ is the growth rate of linear
perturbations, $x_{HI}$ is the mean neutral hydrogen fraction and
$b_{HI}$ is the HI bias. Note that $A_c= C_F C_T$ is the amplitude of
the cross-correlation power spectrum, whereas $A_F=C_F^2$ and
$A_T=C_T^2$ refer to amplitudes of the respective auto-correlation
power spectra.

DLA observations \citep{prochaska2009,noterdaeme2012,zafar2013} in the
redshift range of our interest give a measurement of $\Omega_{HI}
\approx 10^{-3}$ which corresponds to $x_{HI}=0.02$. Semi-numerical
simulations \citep{bagla2010,guha-sarkar2012,villaescusa2014modelling}
of the expected 21-cm signal have been found to be consistent with a
linear, scale-independent HI bias at scales $k \leq 1 \, {\rm
  Mpc}^{-1}$. We have used a recent fit to the simulated $b_{HI}(z)$
\citet{sarkar2016} for the present analysis.  The value of $f(\Omega)$
was calculated using the $\Lambda$CDM cosmological parameters
\citep{planck2014}.

So far we have considered the fluctuations in the transmitted flux of
Lyman-$\alpha$ forest $\delta_F(\vec{\theta}, \Delta z_n)$ to be a
continuous field that is defined at each $\vec{\theta}$ in the FoV. In
reality, this is only observed along a few, discrete lines of sight to
background quasars.  We incorporate this through a sampling function
\begin{equation}
\rho(\vec{\theta}, \Delta z_n) = \frac{\sum_a \, \delta^2
  (\vec{\theta} - \vec{\theta}_a)}{\bar{n}_Q}
\end{equation}
where the sum runs over the background quasars in the FoV, and
$\bar{n}_Q$ is the angular number density of background quasars.
Accounting for the discrete sampling and also the pixel noise in the
observation, the observed dimensionless fluctuation of the
Lyman-$\alpha$ forest transmitted flux can be expressed as
\begin{equation}
\delta_{Fo}(\vec{\theta}, \Delta z_n) = \rho(\vec{\theta}, \Delta z_n)
\, [ \delta_{F}(\vec{\theta}, \Delta z_n) + \delta_{FN}(\vec{\theta},
  \Delta z_n) ]
\label{eq:deltaFo}
\end{equation}
where $\delta_{FN}(\vec{\theta}, \Delta z_n)$ is the contribution from
the pixel noise.

 We model the pixel noise $\delta_{FN}(\vec{\theta}, \Delta z_n)$ as a
 Gaussian random variable with the assumption that the noise in
 different pixels $(\vec{\theta}_a, \Delta z_n)$ are uncorrelated, ie.

\begin{equation}
\langle \delta_{FN}(\vec{\theta}_a, \Delta z_m) \,
\delta^{*}_{FN}(\vec{\theta}_b, \Delta z_n) \rangle = \sigma_{FN}^2 \,
\delta_{ab} \, \delta_{mn}
\end{equation}
where $\sigma_{FN}^2$ is the variance of the pixel noise. We can now
write the observed Lyman-$\alpha$ forest power spectrum,
$\mathcal{P}_{FFo} (\vec{U}, \tau_m)$ as \citep{guha-sarkar2013}
 \begin{equation}
\mathcal{P}_{FFo} (\vec{U}, \tau_m) = \mathcal{P}_{FF} (\vec{U},
\tau_m) + (\bar{n}_Q)^{-1} \, [p_{1D}(\tau_m) + \sigma_{FN}^2 \,
  \Delta z_c]
\label{eq:ff}
\end{equation}
where $p_{1D}(\tau_m)$ is the Lyman-$\alpha$ forest 1D power spectrum
measured in various observations. \citep{borde2013,Palanque2013}.  We
note that the term $(\bar{n}_Q)^{-1} \, [...]$ in eq.~(\ref{eq:ff})
arises due the finite quasar sampling, and this term tends to zero in
the continuum limit $(\bar{n}_Q \rightarrow \infty)$.

\section{Estimates for OWFA}
\label{sec:3}

OWFA PII is a N-S linear array of $N_A=264$ antennas each of which has
a rectangular aperture of dimension $b \times d$, where $b=30 \, {\rm
  m}$ is the width of the ORT parabolic cylinder and $d = 1.92 \, {\rm
  m}$ along the length of the cylinder. The antennas are arranged end
to end along the length of the cylinder resulting in an antenna
separation of $d = 1.92 \, {\rm m}$ between the centers of two
successive antennas. Using a coordinate system with the $x$ and $y$
axes along the length and width of the array respectively, the
aperture power pattern can be written as, \citep{ali-bharadwaj2014}
\begin{equation}
\tilde{a}(\vec{U}, \nu) = \left( \frac{\lambda^2}{bd} \right) \,
\Lambda \left( \frac{U_x \lambda}{d} \right) \, \Lambda \left(
\frac{U_y \lambda}{b} \right) \,.
\end{equation}
where $\Lambda(x)$ is the triangular function defined as,
$\Lambda(x)=1$ for $|x|<1$ and zero elsewhere.  Here
$\tilde{a}(\vec{U}, \nu)$ is the Fourier transform of the OWFA PII
primary beam pattern, $A(\vec{\theta}, \nu) = {\rm sinc}^2 (\pi d
\theta_x/\lambda) {\rm sinc}^2(\pi b \theta_y/\lambda)$ with
corresponds to a FoV of
$(\lambda/d,\lambda/b)=(27^{\circ},1.75^{\circ})$.

The one-dimensional configuration of the antennas allows us to write
the baselines $\vec{U}_n$ as
\begin{equation}
\vec{U}_n = n \left( \frac{d}{\lambda} \right) \hat{i} \hspace{2.5cm}
(1 \leq n \leq N_A-1) \,.
\end{equation}
 The OWFA baselines are highly redundant, and any given baseline
 $\vec{U}_n$ has a redundancy factor of $(N_A-n)$. The visibility
 $\mathcal{V}_T(\vec{U}_a, \nu_n)$ measured at any baseline
 $\vec{U}_n$ and frequency channel $\nu_n$ is related to $\delta
 I(\vec{\theta}, \nu_n)$ the specific intensity fluctuations on the
 sky as
\begin{equation}
\mathcal{V}_T(\vec{U}_a, \nu_n) = \int \, A(\vec{\theta},\nu_n) \,
\delta I(\vec{\theta}, \nu_n) \, e^{2 \pi i \vec{U}_a \cdot \theta}
d^2\theta \, + \mathcal{N}(\vec{U}_a, \nu_n), \,.
\label{eq:vs1}
\end{equation}
where $\mathcal{N}(\vec{U}_a, \nu_n),$ is the additive noise which is
inherent to radio-interferometric observations. We further assume that
the noise in different baselines and frequency channels is
uncorrelated, and the real and imaginary components of
$\mathcal{N}(\vec{U}_a, \nu_n)$ both have zero mean and standard
deviation
\begin{equation}
\sigma_T = \frac{\sqrt{2} k_B T_{sys}}{ \eta A \sqrt{\Delta \nu_c t}}
\label{eq:vis2}
\end{equation}
where $T_{sys}$ is the system temperature, $\eta$ is the aperture
efficiency, $A=b \times d$ is the aperture area, $\Delta \nu_c$ is the
frequency channel width which also corresponds to $\Delta z_c$ and $t$
is the observation time. Note that we have assumes a single
polarization (which is the case for OWFA) and not incorporated the
baseline redundancy in eq. (\ref{eq:vis2}).

For the redshifted 21-cm signal we replace $\nu_n$ with $\Delta z_n$
using a the conversion $\nu = 1420/(1+z)$, and use $\delta
I(\vec{\theta}, \nu) = Q_\nu \, \delta_T(\vec{\theta}, \nu)$ where
$Q_\nu=2 k_B/\lambda^2$ is the conversion factor from brightness
temperature to specific intensity in the Rayleigh-Jeans approximation.
The baselines $\vec{U}$, the primary beam pattern $A(\vec{\theta},
\nu)$ and $Q_\nu$ all change with frequency. For the present analysis
we assume that this variation is relatively small ($ \sim 15-20 \%$ as
shown for OWFA PI in\citet{sarkar2017analytical}) across the observing
bandwidth and we hold these quantities fixed at the values
corresponding to central frequency $\nu_c$.

It is convenient to decompose the visibilities $\mathcal{V}_T(\vec{U},
\Delta z_n)$ into delay channels \citep{morales2005}, using
\begin{equation}
v_{T}(\vec{U}, \tau_m) = \Delta z_c \sum_{n} e^{2 \pi i \tau_m (n
  \Delta z_c)} \, \mathcal{V}_{T}(\vec{U}, \Delta z_n)
\label{eq:tau}
\end{equation}
whereby we can express $v_{T}(\vec{U}, \tau_m)$ in terms of
$\Delta_T(\vec{U}^{'},\tau_m)$ (eq.~\ref{eq:a1}) as
\begin{equation}
v_T(\vec{U}_a, \tau_m) = Q \int \, \tilde{a}(\vec{U}_a-\vec{U}^{'}) \,
\Delta_T(\vec{U}^{'},\tau_m) \, d^2 U^{'} + n(\vec{U}^{'},\tau_m) \,.
\label{eq:VHI}
\end{equation}
and the noise contribution now has the property
\begin{equation}
\langle n^{*}(\vec{U}_a, \tau_n) n(\vec{U}_b,
\tau_m) \rangle = 2 B \Delta z_c \, \sigma_T^2 \, \delta_{a,b} \,
\delta_{n, m} \,.
\label{eq:noise}
\end{equation}

A Lyman-$\alpha$ forest survey typically covers a large area of the
sky (e.g. BOSS covers an area of 6373.2 deg$^{2}$ for DR10 of
SDSS\citet{ahn2014tenth}) as compared to the OWFA FoV. Considering
OWFA observations in a particular pointing direction, the 32-cm
visibility signal will be maximally correlated with the Lyman-$\alpha$
forest signal when they both originate from exactly the same region,
and the correlation will fall off ($\sim 0$) if the regions from where
the two signals originate have no overlap.  An uncorrelated component
in the Lyman-$\alpha$ forest will contribute to the variance without
contributing to the cross-correlation signal, thereby degrading the
signal to noise ratio. For the cross-correlation signal it is most
advantageous to exactly match the spatial regions which are probed by
the 21-cm and the Lyman-$\alpha$ respectively. We address this by
multiplying the Lyman-$\alpha$ forest data $\delta_{Fo}(\vec{\theta},
\Delta z_n)$ with the OWFA primary beam pattern
$A(\vec{\theta},\nu_c)$ and restricting the redshift band $B$ to
exactly match the OWFA coverage. In analogy with th 21-cm signal
visibilities, we also decompose the Lyman-$\alpha$ forest signal into
visibilities (eq.~\ref{eq:vs1}) at the OWFA baselines.  It is useful
to express these Lyman-$\alpha$ visibilties as
 \begin{equation}
v_F(\vec{U}, \tau_n) = \int \, \tilde{a}(\vec{U}-\vec{U}^{'}) \,
\Delta_{Fo}(\vec{U}^{'}, \tau_n) \, d^2 U^{'} \,.
\label{eq:visa}
\end{equation}
Note that $\Delta_{Fo}(\vec{U}^{'}, \tau_n)$ includes the pixel noise
(eq. \ref{eq:deltaFo}) and there is no other additive noise term in
eq.~(\ref{eq:visa}).

We use the visibility correlation
\begin{equation}
C_{ab}(m) = \langle v^{*}_{T}(\vec{U}_a, \tau_m) v_{F}(\vec{U}_b,
\tau_m) \rangle
\label{eq:cov1}
\end{equation}
to quantify the cross-correlation signal predictions for OWFA, and
using equations (\ref{eq:pk}), (\ref{eq:VHI}) and (\ref{eq:visa}) we
obtain
\begin{equation}
C_{a\,b}(m)=Q \, B \, \int \tilde{a}(\vec{U}_a-\vec{U}^{'}) \,
\tilde{a}(\vec{U}_b-\vec{U}^{'}) \, P_{FT}(\vec{U}^{'}, \tau_m) \, d^2
U^{'} \,.
\label{eq:cov2}
\end{equation}
We note that the signal at different $\tau_m$ values are uncorrelated
(eq.~\ref{eq:pk}), and it is not necessary to consider such
correlations.  The visibilities measured at OWFA are correlated only
for the same baseline ($a=b$) and adjacent baselines ($a=b \pm
1$). Further the correlation for adjacent baselines is roughly one
fourth the correlation at the same baseline \citep{ali-bharadwaj2014},
and all other possible correlations are zero.  It is convenient to
introduce a more compact notation using $\mathbf{P}_a(m)$ with $1 \le
a \le 2 N_A - 3$ to denote the non-zero elements of the visibility
correlation $C_{ab}(m)$. The first $N_A-1$ elements of
$\mathbf{P}_a(m)$ denote the correlations at the same baseline
($\mathbf{P}_a(m)=C_{ab}(m)$ with $b=a$), and the subsequent $N_A-2$
elements of $\mathbf{P}_a(m)$ denote the correlations at the adjacent
baselines ($\mathbf{P}_a(m)=C_{ab}(m)$ with $b=a+1$).  The error
covariance for $\mathbf{P}_a$ can be calculated using
\begin{align}
 \langle &\mathbf{\Delta P}_{a_1}(m) \mathbf{\Delta P}_{a_2}(m)
 \rangle = \frac{1}{2} \, \left[ C_{{a_1}{a_2}}(m) \,
   C_{{b_1}{b_2}}(m) + C_{{a_1}{b_2}}(m) \, C_{{a_2}{b_1}}(m) \right]
 +
\label{eq:covar}
\\ & \frac{1}{8} \, \left[ F_{{a_1}{a_2}}(m) \, T_{{b_1}{b_2}}(m) +
  F_{{b_1}{b_2}}(m) \, T_{{a_1}{a_2}}(m) +F_{{a_1}{b_2}}(m) \,
  T_{{a_2}{b_1}}(m) + F_{{a_2}{b_1}}(m) T_{{a_1}{b_2}}(m) \right] \,.
\label{eq:errcov}
\end{align}
where $b_1$ and $b_2$ are indices associated with $a_1$ and $a_2$
respectively. We find that the error covariance in the above equation
(eq.~\ref{eq:errcov}) depends on terms of two different types, the
first being the product of the cross correlations signals
eg. $C_{{a_1}{b_2}}(m) \, C_{{a_2}{b_1}}(m)$ and the second being the
product of the auto correlation of the Lyman-$\alpha$ forest and the
HI 21-cm signal, eg. $F_{{b_1}{b_2}}(m) \, T_{{a_1}{a_2}}(m)$. We
calculate the auto correlation signals of the HI 21-cm signal and the
Lyman-$\alpha$ forest respectively using

\begin{align}
T_{ab}(m)=&B Q^2 \int \tilde{a}(\vec{U}_a-\vec{U}^{'}) \,
\tilde{a}(\vec{U}_b-\vec{U}^{'}) \, P_{TT}(\vec{U}^{'}, \tau_m) \, d^2
U^{'} + \nonumber \\ &2 B \, \sigma_T^2 \, \Delta z_c \, (N_A-a)^{-1}
\, \delta_{a,b}
 \label{eq:T1}
\end{align}
and
\begin{equation}
F_{ab}(m)=B \int \tilde{a}(\vec{U}_a-\vec{U}^{'}) \,
\tilde{a}(\vec{U}_b-\vec{U}^{'}) \, P_{FFo}(\vec{U}^{'}, \tau_m) \,
d^2 U^{'}
\label{eq:F1}
\end{equation}
where the term $(N_A-a)^{-1}$ in eq~(\ref{eq:T1}) arises from the OWFA
baseline redundancy.

We have used the Fisher matrix
\begin{equation}
\mathbf{F}_{pq} = \sum_{m} \, \mathbf{P}_{{a_1},p}(m) \, [\langle
  \mathbf{\Delta P}_{a_1}(m) \mathbf{\Delta P}_{a_2}(m) \rangle]^{-1}
\, \mathbf{P}_{{a_2},q}(m),
\label{eq:fisher}
\end{equation}
to estimate the accuracy for parameter estimation using OWFA
measurements of the 21-cm and Lyman-$\alpha$ cross-correlation signal,
here the indices $p$ and $q$ refer to the different parameters whose
values we wish to estimate from the observations.  For most of our
analysis we have considered three parameters $q_1={\rm ln}(A_c)$,
$q_2={\rm ln}(\beta_T)$ and $q_3={\rm ln}(\beta_F)$ which correspond
to the amplitude of the cross-correlation signal, the redshift space
distortion parameter for the 21-cm signal and the redshift space
distortion parameter for the Lyman-$\alpha$ forest respectively.  The
elements $\delta A_c/A_c=1/\sqrt{\mathbf{F}_{11}}$, $\delta
\beta_T/\beta_T=1/\sqrt{\mathbf{F}_{22}}$ and $\delta \beta_F/\beta_F
=1/\sqrt{\mathbf{F}_{33}}$ provide estimates of the conditional
fractional errors in the respective parameters.  We have inverted the
Fisher matrix to calculate the marginalized errors reported
subsequently in this paper.

The dependence on the HI 21-cm observation time $t$ enters into the
Fisher matrix (eq.~\ref{eq:fisher}) through the error covariance
(eq.~\ref{eq:errcov}). For small observation times, the noise (second
term in eq.~\ref{eq:T1}) dominates over the HI 21-cm signal and we
have $T_{ab}(m) \propto 1/t$. Further, the terms of the form
$T_{ab}(m)F_{ab}(m)$ dominate the error covariance $\langle
\mathbf{\Delta P}_{a_1}(m) \mathbf{\Delta P}_{a_2}(m) \rangle$ whereby
we expect the Fisher matrix to increase linearly with $t$,
ie. $\mathbf{F}_{pq}\propto t$. For large observation times, the noise
(second) term in eq.~(\ref{eq:T1}) is considerably below the HI 21-cm
signal and we expect $T_{ab}(m)$ and also the Fisher matrix to
saturate at a value which is independent of the observation time.

Apart from the observation time, the Fisher matrix
(eq.~\ref{eq:fisher}) also depends on the quasar number density
$\bar{n}_Q$ and the pixel noise $\sigma_{FN}$ through
$\mathcal{P}_{FFo}$ (eq.~\ref{eq:ff}) which affects the error
covariance (eq.~\ref{eq:errcov}) through $F_{ab}(m)$
(eq.~\ref{eq:F1}).  We see that it is possible to decrease the error
covariance by increasing the quasar number density $\bar{n}_Q$. It is
also possible to decrease the error covariance by reducing
$\sigma_{FN}$.  We expect the error covariance and the Fisher matrix
to saturate in the continuum limit, $\bar{n}_Q \rightarrow \infty$.
We finally note that the limits $\bar{n}_Q \rightarrow \infty$ and $t
\rightarrow \infty$ corresponds to the cosmic variance which sets the
lower and upper limits for the error covariance and the Fisher matrix
respectively. For the present analysis consider $\bar{n}_Q(z)$ and
$\sigma_{FN}$ to be fixed as given by SDSS DR-14 \citep{SDSS14}, and
discuss these in somewhat more detail in the subsequent section.

The analysis of this paper is mainly focused on the HI 21-cm and
Lyman-$\alpha$ forest cross-correlation signal. As discussed earlier, we
expect the Fisher matrix for the cross-correlation to scale as
$\mathbf{F}_{pq}\propto t$ for small observation times and
subsequently saturate for large observation times. For comparison we
also consider the HI 21-cm auto-correlation signal. In contrast to the
error covariance for the cross-correlation (eq.~\ref{eq:errcov}), the
error covariance for the 21-cm auto-correlation only has terms of the
form $ \langle \mathbf{\Delta P}_{a_1}(m) \mathbf{\Delta P}_{a_2}(m)
\rangle \sim \propto T_{{a_1}{b_2}}(m) \, T_{{Pa_2}{b_1}}(m)$ (eq. 7
of \citet{bharadwaj2015fisher}). We therefore expect the Fisher matrix
for the auto-correlation to scale as $\mathbf{F}_{pq}\propto t^2$ for
small observation times and subsequently saturate at the cosmic
variance limit for large observation times. It is useful to note that
the Fisher matrix for the auto-correlation signal increases more
rapidly than that for the cross-correlation signal as the observation
time is increased.  It also follows that for small observation times
we expect the errors in parameter estimation to scale as $\delta q
\propto 1/\sqrt{t}$ and $\delta q \propto 1/t$ for the
cross-correlation and auto-correlation signals respectively. In both
cases the errors saturate at large $t$.

\section{Observational Considerations}

We have carried out our present analysis considering OWFA PII which
operates at a central frequency $\nu_c = 326.5 \, {\rm MHz}$ that
corresponds to HI at redshift $z_c = 3.35$. Although the actual
bandwidth of the OWFA PII is $ 39 \, {\rm MHz}$, for the present
analysis we have used a bandwidth of $B = 30 \, {\rm MHz}$ that gives
access to observation in the range $3.15 \leq z \leq 3.55$.  It is
worthwhile and interesting to consider the possibility of modifying
the antenna system to increase the OWFA bandwidth.  In order to assess
the improvement that would be achieved by increasing the bandwidth, we
have also performed the entire analysis considering a larger bandwidth
of $B = 60 \, {\rm MHz}$.

We have used the quasar number distribution from the DR14 of SDSS
\citep{SDSS14} which gives the binned quasar numbers $\Delta N(z)$
over different redshifts in the range $0 \leq z \leq 4$, with a bin of
width $\Delta z_{{\rm bin}} = 0.04$. We see that the quasar
distribution peaks at $z = 2.25$, with $\Delta N(z) \sim 30000$, and
falls off as we move away from the peak. We further see that at $z
\geq 3$, $\Delta N(z)$ goes below one-tenth of the value at $z =
2.25$.  Note that SDSS DR14 has a total angular coverage of $14,555 \,
{\rm deg}^2$ \footnote{http://www.sdss.org/dr14/scope/}.

The number of quasars available for estimating the cross correlation
signal in the OWFA band centered at the redshift, $z_c = 3.35$ is
quite low. We have explored the possibility to increase the number of
quasars by considering observations at a redshift nearer to the peak
of the quasar distribution.  For the purpose of the present analysis,
we have considered the possibility of modifying the OWFA system for
observation at redshifts $z_c = 3.05 \, {\rm and} \, z_c = 2.55$ that
correspond to HI 21-cm observation at frequencies $\nu_c = 350 \, {\rm
  MHz} \, {\rm and} \, \nu_c = 400 \, {\rm MHz}$ respectively. The
number of quasars increases by a factor of $\sim 3$ and $\sim 5$
respectively as compared to that at redshift $z_c = 3.35$.

Given a quasar at redshift $z_Q$, for the cross-correlation analysis
we have excluded the region $10,000 \, {\rm km \, s^{-1}}$ blue-ward
of the Lyman-$\alpha$ emission peak of the quasar because of the
quasar proximity effect \citep{bajtlik1988quasar,weymann1981quasar}.
We have further only considered the part of the spectrum beyond $1,000
\, {\rm km \, s^{-1}}$ red-ward of the Lyman-$\beta$ line or the O-VI
lines to avoid any possibility of confusing the Lyman-$\alpha$ forest
with the Lyman-$\beta$ or O-VI absorption lines in the quasar
spectrum. For the given quasar, this restricts the redshift range
across which the quasar spectrum can be used for the cross-correlation
analysis. The overlap between this redshift range and the OWFA 21-cm
band can vary depending on the values of $z_Q$, $z_c$ and $B$.  We
have accounted for both complete and partial overlap in estimating the
mean quasar number density $\bar{n}_Q(z_c)$ available for the
cross-correlation analysis. The actual signal to noise ratio (SNR)
varies from quasar to quasar, and this can be as large as SNR=10 for
the bright quasars.  To keep
the analysis simple we have assumed an uniform value ${\rm SNR} = 5$
which implies $\sigma_{FN}=0.2$ in eq.~(\ref{eq:ff}).

The discussion till now is restricted to 21-cm observations in a
single pointing direction. As mentioned earlier, the OWFA FoV is much
smaller than the area covered by Lyman-$\alpha$ surveys like BOSS, and it
is worthwhile to also consider the possibility of extending the
analysis to a situation where 21-cm observations are carried out in
$N_p$ different pointing directions. In the present work we have
assumed that we obtain independent cross-correlation signals from each
different pointing direction whereby the Fisher matrix for the
combined observation is $N_p$ times the Fisher matrix for a single
pointing direction. The results from our Fisher matrix analysis are
presented in the section to follow.

\section{Results}

\begin{figure}
\begin{center} 
\psfrag{A305}[c][c][0.5][0]{$B=30$ {\rm MHz} \quad \quad \quad}
\psfrag{B305}[c][c][0.5][0]{$B=30$ {\rm MHz} \quad \quad \quad}
\psfrag{A605}[c][c][0.5][0]{$B=60$ {\rm MHz} \quad \quad \quad}
\psfrag{B605}[c][c][0.5][0]{$B=60$ {\rm MHz} \quad \quad \quad}
\psfrag{zA}[c][c][0.7][0]{\quad \quad \quad $z_c=3.35$}
\psfrag{zB}[c][c][0.7][0]{\quad \quad \quad \quad $z_c=3.05$}
\psfrag{zC}[c][c][0.7][0]{\quad \quad \quad \quad $z_c=2.55$}
\psfrag{k1}[c][c][1.0][0]{SNR} \psfrag{k2}[c][c][1][0]{$t$ (hours)}
\psfrag{ta}[c][c][0.8][0]{\qquad auto}
\psfrag{tc}[c][c][0.8][0]{\qquad cross}
\psfrag{sa}[c][c][0.8][0]{$t$ \quad}
\psfrag{sc}[c][c][0.8][0]{$t^{0.5}$}
\vskip.2cm
\centerline{{\includegraphics[scale=0.4,angle=-90]{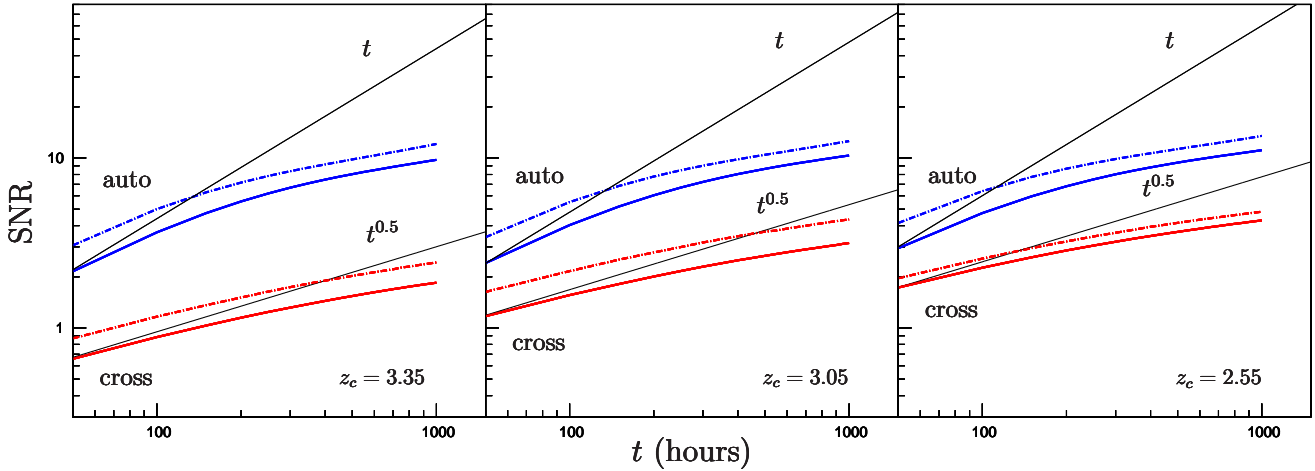}}}
\caption{Shows the plot of the predicted SNR for measuring $A_c$ (red
  lines) and $C_T$ (black lines) versus the observing time $t$ for the
  different redshifts and bandwidths considered for our analysis. The
  solid black lines in each figure shows the expected behaviour of the
  SNR for measuring $A_c$ and $C_T$ with $t$ respectively, ie. ${\rm
    SNR} \propto t^{0.5}$ and ${\rm SNR} \propto t$, in the limit of
  small observing times.}
\label{fig:snr_mnd}
\end{center}
\end{figure}

We begin our analysis by considering the prospects of measuring the
amplitude of the cross correlation signal $A_c$ using OWFA PII.
Figure \ref{fig:snr_mnd} shows how the predicted signal-to-noise ratio
(SNR) for measuring $A_c$ varies with the observing time $t$. We have
calculated the SNR for measuring $A_c$ using ${\rm SNR} =
1/\sqrt{[F^{-1}]_{11}}$ which is marginalized over $\beta_T$ and
$\beta_F$ the two other parameters in our analysis. We first consider
the left panel of the Figure \ref{fig:snr_mnd} which shows the results
at $z=3.35$ where OWFA is expected to operate at present.  As
discussed in Section \ref{sec:3}, at small observation times we expect
the SNR to increase as, ${\rm SNR} \propto t^{0.5}$.  We see that for
$t $ in the range $ 50-100$ hrs the SNR grows as ${\rm SNR} \propto
t^{0.43}$ which is somewhat slower compared to ${\rm SNR} \propto
t^{0.5}$.  We may expect the ${\rm SNR} \propto t^{0.5}$ regime to be
restricted to observing times $t < 50$ hrs. We see that for $B = 30 \,
{\rm MHz}$ the SNR increases to $\sim 1$ for $t \sim 200$ hrs. The SNR
increases very slowly beyond this, and it only increases by a factor
of $\sim 1.5$ for $t \sim 1000$ hrs. Considering a larger bandwidth of
$B = 60 \, {\rm MHz}$, we see that the SNR increases by a factor of
$\sim 1.3$ as compared to $B = 30 \, {\rm MHz}$.

As mentioned earlier, the number of quasars available for the
cross-correlation analysis peaks around $z \sim 2.25$. In order to see
if it is possible to improve the prospects of a measurement by moving
closer to the peak, we have considered observations centered at two
other redshifts namely $z_c=3.05$ and $2.55$ for which the results are
shown in the middle and right panels respectively. Considering $B = 30
\, {\rm MHz}$ we see that it is possible to achieve ${\rm SNR} \sim 2$
and $\sim 3$ with $t \sim 200$ hrs observations at $z_c=3.05$ and
$2.55$ respectively.  Considering $B=60 \, {\rm MHz}$, the SNR values
increase by factors of $1.4$ and $1.1$ at $z_c=3.05$ and $2.55$
respectively. In all cases, the SNR increases very slowly for
observation times $t > 200$ hrs, and it is not very meaningful to
consider deeper observation in a single FoV. Based on this, we
consider a situation where $N_p=25$ independent fields-of-view are
observed for $200$ hrs each. The predicted SNR values for measuring
$A_c$ are summarized in the Table \ref{tab:snr_mnd}. We see that a
$5-\sigma$ detection is possible in all the cases considered here. We
have SNR~$\sim 6$ at $z_c=3.35$ with $B=30 \, {\rm MHz}$. The SNR
increases to $\sim 15$ at $z_c=2.55$ which is closer to the peak of
the quasar distribution, the SNR also increases by a factor of $1.1$
to $1.5$ if the bandwidth is doubled.

For comparison, we have also shown the SNR for measuring the HI 21-cm
auto-correlation signal $C_T$ (blue lines). We see
that in this case  the SNR varies rather rapidly with $t$ and
subsequently saturates at a larger $t$ as compared to the SNR for
measuring $A_c$. For $B = 30 \, {\rm MHz}$, it is possible to have a
measurement with ${\rm SNR} \sim 5$ for $t \sim 150$ hrs at $z_c =
3.35$. We further see that the SNR increases by a factor of $\sim 1.1$
and $\sim 1.2$ at $z_c = 3.05$ and $2.55$ respectively. The SNR
increases by a factor of $1.4$ for all the redshifts considered here
if the bandwidth is doubled. 

\begin{table}
\begin{center}
\caption{Predicted SNR for measuring $A_c$ with an observation time of
  $200$ hrs each in 25 independent fields-of-view for the different
  redshifts and bandwidths considered here.}
\vspace{.2in}
\label{tab:snr_mnd}
\begin{tabular}[scale=.6]{|c|c|c|c|c|c|c|}
\hline z & 3.35 & 3.35 & 3.05 & 3.05& 2.55 & 2.55 \\ 
\hline B (MHz) & 30 & 60 & 30 & 60 & 30 & 60 \\ 
\hline SNR & $\sim 6$ & $\sim 9$ & $\sim 10$ & $\sim 14$ & $\sim 15$ & $\sim 17$ \\ 
\hline
\end{tabular}
\end{center}
\end{table}

\begin{figure}
\begin{center}
\vskip.2cm \psfrag{k2}[c][c][0.65][0]{$\Delta[{\rm ln}(\beta_F^{0.83}
    \beta_T^{0.55})]$} \psfrag{k1}[c][c][0.65][0]{$\Delta[{\rm
      ln}(\beta_F^{-0.55} \beta_T^{0.83})]$}
\psfrag{k4}[c][c][0.65][0]{$\Delta[{\rm ln}(\beta_F^{0.82}
    \beta_T^{0.57})]$} \psfrag{k3}[c][c][0.65][0]{$\Delta[{\rm
      ln}(\beta_F^{-0.57} \beta_T^{0.82})]$}
\psfrag{k6}[c][c][0.65][0]{$\Delta[{\rm ln}(\beta_F^{0.84}
    \beta_T^{0.54})]$} \psfrag{k5}[c][c][0.65][0]{$\Delta[{\rm
      ln}(\beta_F^{-0.54} \beta_T^{0.84})]$}
\psfrag{zA}[c][c][0.6][0]{\quad $z_c = 3.35$}
\psfrag{zB}[c][c][0.6][0]{$z_c = 3.05$} \psfrag{zC}[c][c][0.6][0]{$z_c
  = 2.55$} \centerline{{\includegraphics[scale =.45]{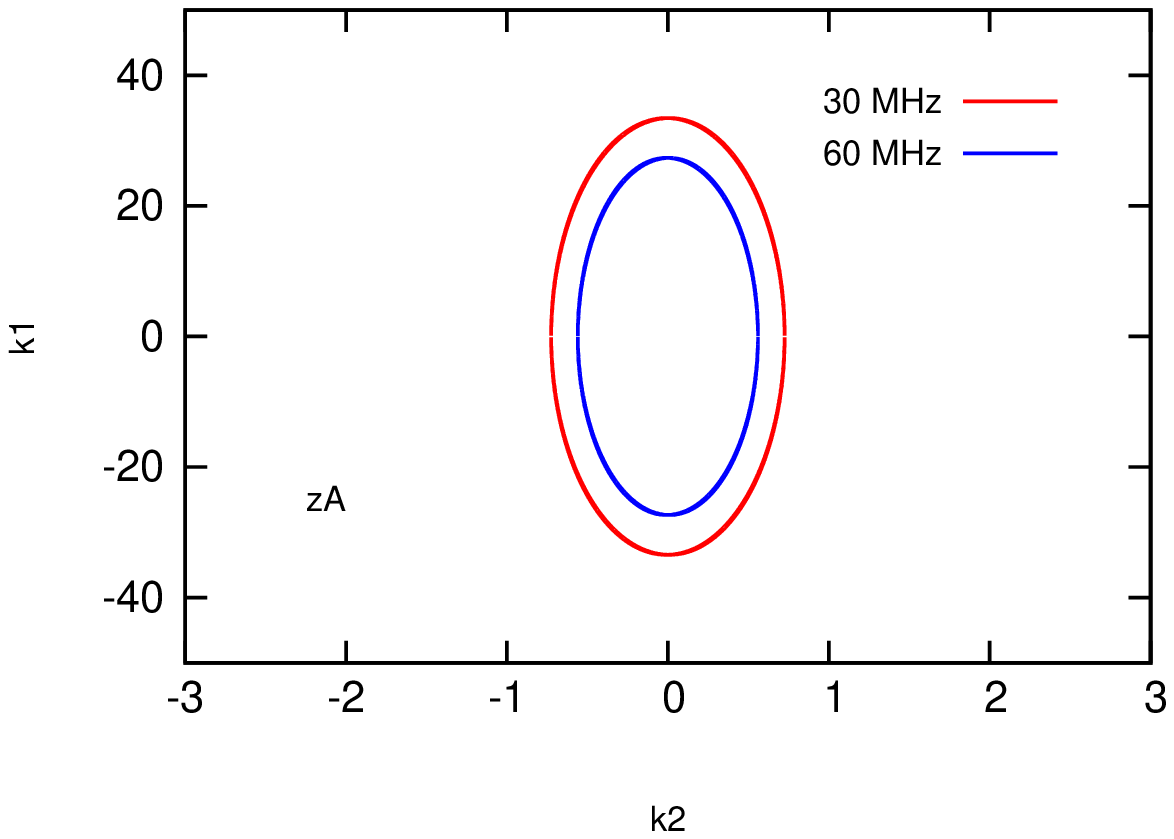}}
  { \includegraphics[scale =.45]{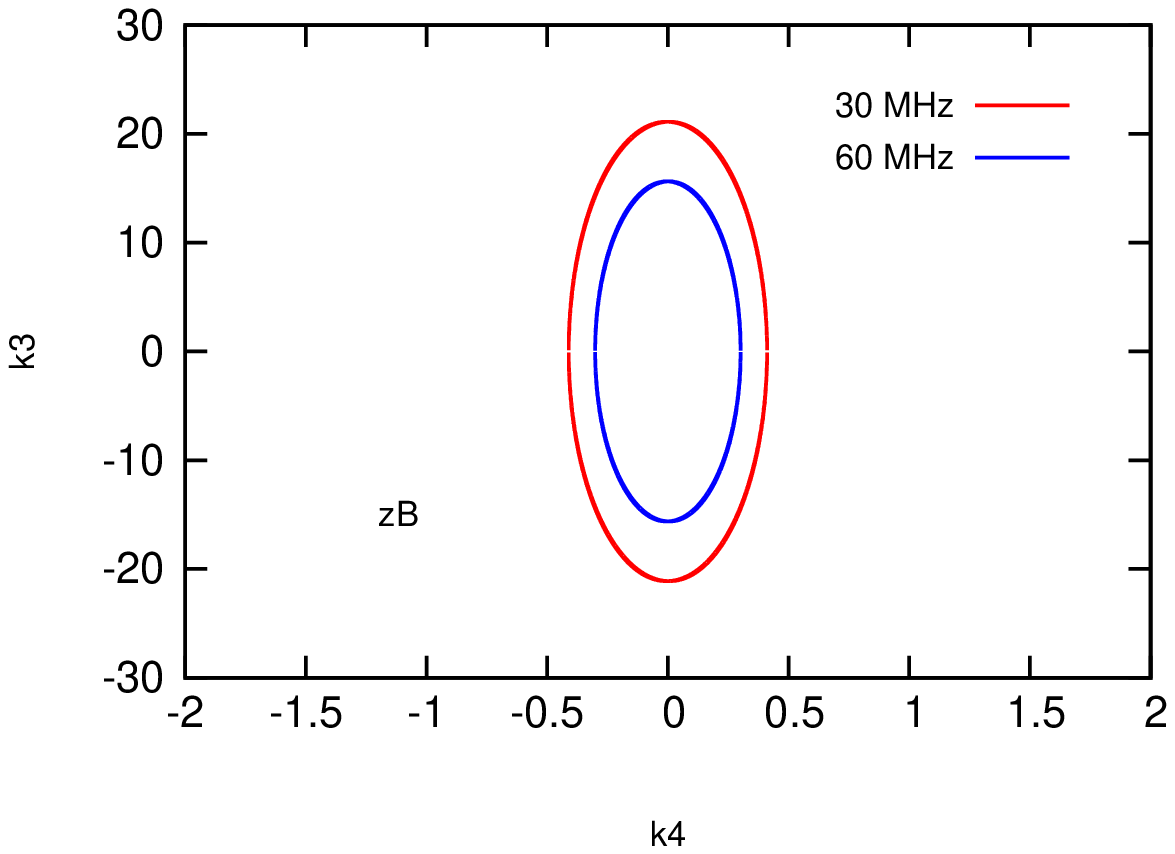}} {
    \includegraphics[scale =.45]{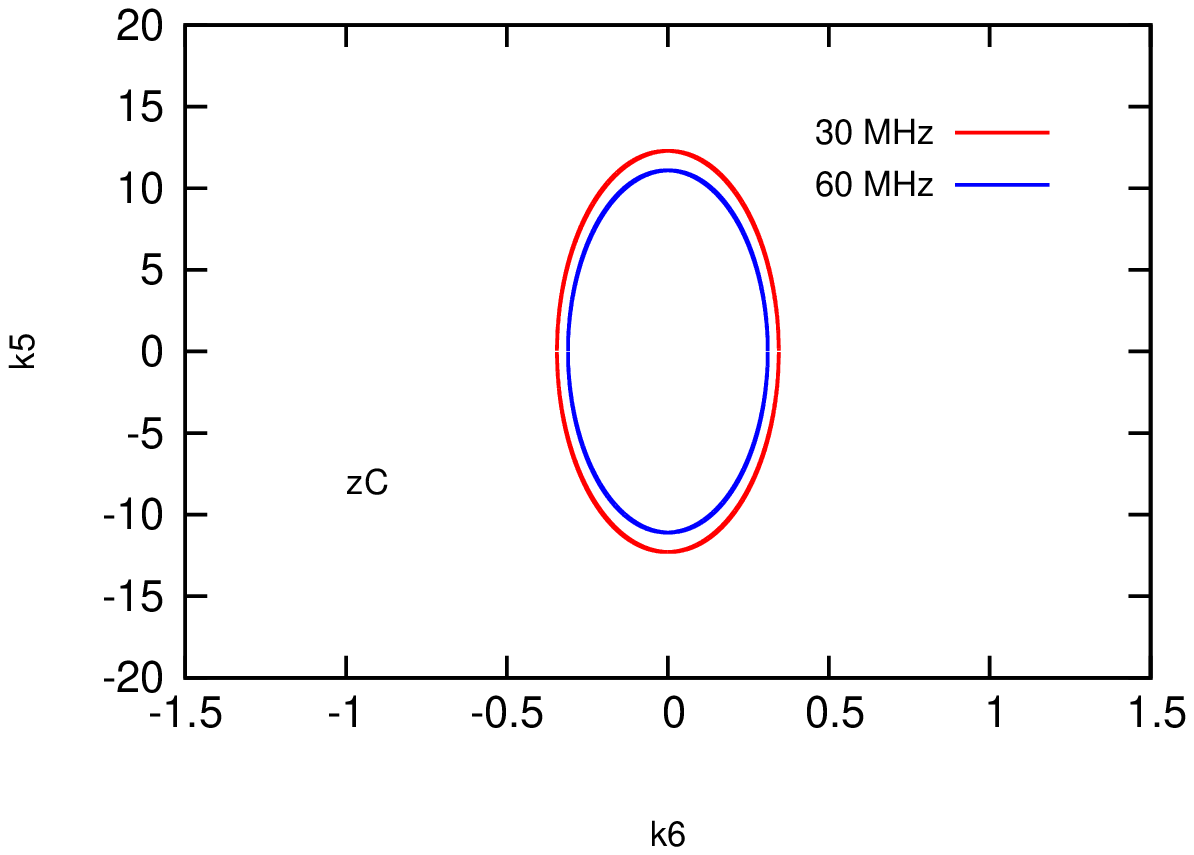}}}
\caption{Shows the errors on the joint measurement of the parameters
  $q^{ab} = {\rm ln}(\beta_F^a \beta_T^b)$, with an observing time of
  $200$ hrs each in $N_p = 25$ independent fields-of-view for the
  different redshifts and bandwidths considered in our analysis.}
\label{fig:fisherjdgnu}
\end{center}
\end{figure}

We have next  considered the prospects of the joint measurement of the
parameters $\beta_T$ and $\beta_F$ marginalizing over $A_c$.  We have
considered an observing time of $200$ hrs each in $N_p=25$ independent
fields-of-view. We find that the relative errors in $\beta_T$ and
$\beta_F$ are large and are highly anti-correlated. We have identified
the combinations $q_1=(\beta_F^{0.83}\beta_T^{0.55})$ and
$q_2=(\beta_F^{-0.55} \beta_T^{0.83})$ for which the errors are
uncorrelated at $z_c=3.35$. Note that the values of the exponents
change slightly with $z_c$. For $z_c = 3.35$ (left panel of Figure
\ref{fig:fisherjdgnu}) with $B = 30 \, {\rm MHz}$ we have relative
errors of $\sim 1$ and $\sim 40$ for $q_1$ and $q_2$ respectively.
The errors on $q_1$ and $q_2$ respectively decrease by factors of $2$
and $1.3$ for $B = 60 \, {\rm MHz}$.  Considering $z_c = 3.05$
(central panel) for $B = 30 \, {\rm MHz}$ we have relative errors of
$\sim 0.5$ and $\sim 20$ for $q_1$ and $q_2$ respectively, and these
decrease respectively by factors of $1.6$ and $1.3$ for $B = 60 \,
{\rm MHz}$. Considering $z_c = 2.55$ (right panel) for $B = 30 \,
{\rm MHz}$ we have relative errors of $\sim 0.4$ and $\sim 15$ for
$q_1$ and $q_2$ respectively, and these decrease respectively by
factors of $1.3$ and $1.25$ for $B = 60 \, {\rm MHz}$.

In the previous discussion we find that the prospects of a joint
measurement of the combinations of $\beta_F$ and $\beta_T$ are
quite low, and at best possible 
a $3 \, \sigma$ measurement is  possible at  $z_c = 2.55$ for $B = 60
\, {\rm MHz}$. However, it is 
worthwhile to note that it is possible to measure $\beta_T$  at a
comparatively higher SNR using HI 21-cm observation alone. As can be
seen from Figure \ref{fig:snr_mnd} (blue lines), the SNR for the auto
correlation signal increases more rapidly with $t$ as compared to the
cross-correlation. For the auto-correlation it is more advantageous to
consider deeper observations of $t \sim 1000$ hrs, and for the present
analysis we consider a situation where such observations are carried
out in $N_p = 5$ independent fields of view to measure the value of
$\beta_T$ from the auto-correlation signal.  We find that it is
possible to measure  $\beta_T$ with  errors $\Delta
\beta_T/\beta_T \sim 0.24, \, 0.22, \, 0.18$ respectively at $z_c =
3.35, \, 3.05, \, 2.55$ for $B = 30 \, {\rm MHz}$.  The errors
decrease by a factor of $2.3$ for  $B = 60 \,
{\rm MHz}$. We consider the relative error $\Delta \beta_F/\beta_F$ in
measuring $\beta_F$ given that the measurement of  $\beta_T$ comes
from the HI 21-cm auto-correlation. For $B = 60 \, {\rm  MHz}$  we 
find $\Delta \beta_F/\beta_F \approx 0.6,0.36$ and $0.23$ at 
$z_c = 3.35, 3.05$ and $2.55$ respectively. For $B = 30 \, {\rm  MHz}$ 
these values go down by  factors of $1.45,1.4$ and $1.15$
respectively. We see that it is possible to measure $\beta_F$ with $4
\, \sigma$ and $5 \, \sigma$ confidence with  $B = 30 \, 
   {\rm  MHz}$  and $60 \,   {\rm  MHz}$  respectively by combining
   auto-correlation and cross-correlation observations at $z_c=2.55$.

The analysis so far has assumed that the shape of the 
cross-correlation power spectrum  $\mathcal{P}_{FT}(U,
\tau_m)$  traces the dark matter power spectrum $P(k)$
(eq.~\ref{eq:PFT}).  It is interesting and worthwhile to
consider the possibility of constraining the shape of the
cross-correlation  power spectrum directly from observation. To this
end we have assumed that the values of $\beta_T$ and $\beta_F$ are
known, and we have investigated how precisely it will be possible to
measure $\mathcal{P}_{FT}(k)$ using observations of the
cross-correlation signal. We have binned the $k$ range $0.018 \le k
\le 2.73 \, {\rm Mpc}^{-1}$  which will be probed by OWFA PII into $5$ 
equally spaced logarithmic $k$-bins and we have used the Fisher
matrix technique to determine the SNR with which it would be possible
to measure the amplitude of the cross-correlation power spectrum in
each of these $k$ bins. Note that the $k$ limits change to some extent
when $z_c$ and $B$ are varied, and we have incorporated these in our
estimates. 

\begin{figure}
\begin{center} 
\vskip.2cm \psfrag{A305}[c][c][0.65][0]{$B=30$ {\rm MHz} \quad \quad
  \quad \quad} \psfrag{A605}[c][c][0.65][0]{$B=60$ {\rm MHz} \quad
  \quad \quad \quad} \psfrag{5}[c][c][0.6][0]{5}
\psfrag{10}[c][c][0.6][0]{10} \psfrag{zA}[c][c][0.75][0]{\quad \quad
  $z_c=3.35$} \psfrag{zB}[c][c][0.75][0]{\quad \quad $z_c=3.05$}
\psfrag{zC}[c][c][0.75][0]{\quad \quad $z_c=2.55$}
\psfrag{k1}[c][c][0.8][0]{SNR} \psfrag{k2}[c][c][0.8][0]{$k \, {\rm
    Mpc}^{-1}$}
\centerline{{\includegraphics[scale=0.4,angle=-90]{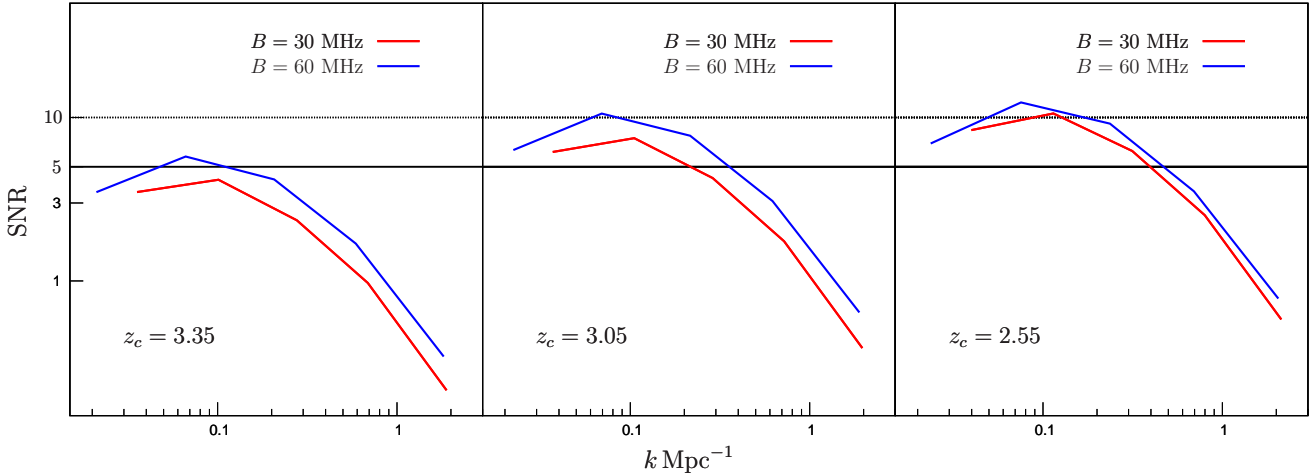}}}
\caption{Shows the plot of SNR for measuring the amplitude of $\mathcal{P}_{FT}(k)$ 
  in different $k$-bins with an observing time of $200$ hrs each in $N_p = 25$
  independent fields-of-view for the different redshifts and
  bandwidths considered here. The lower solid black lines and upper
  dash-dotted black lines correspond to SNR $= 5$ and $10$ respectively.}
\label{fig:snr_bin}
\end{center}
\end{figure}

Figure \ref{fig:snr_bin} shows the predicted SNR for measuring the 
amplitude of the cross-correlation power spectrum  in different  $k$
bins.  We expect the errors in the  small $k$-bins to be  dominated by
cosmic variance as opposed to  the larger $k$-bins where the
errors are predominantly due to  the system noise. We see that 
for $z_c = 3.35$ (left panel) and $B = 30 \, {\rm MHz}$ a ${\rm
  SNR} \ge 3$ measurement is possible in the two smallest $k$ bins,
however it is not possible to achieve  ${\rm SNR} \sim 5$  in 
any of the $k$-bins with  $200$ hrs observations in each of
$N_p=25$ independent fields considered  here. If the 
bandwidth is increased to $B = 60 \, {\rm MHz}$, 
a measurement with  ${\rm SNR} \geq 5$ is possible in a single $k$-bin
which is centered at $k \approx 0.7 \, {\rm Mpc}^{-1}$. 
As  mentioned earlier, we expect the SNR to increase if $z_c$ is tuned
toward the peak of  the quasar distribution at $z \approx 2.25$. 
We find that for $z_c = 3.05$ (central panel) and $B = 30 \, {\rm MHz}$, we can have
a measurement with  ${\rm SNR} \geq 5$ in the two smallest  $k$-bins  
($\le 0.1 \, {\rm Mpc}^{-1}$), and ${\rm SNR} \sim
3$ is possible for the third smallest $k$-bin at $k \approx 0.3 \,
{\rm Mpc}^{-1}$. For $B = 60 \, {\rm MHz}$ we expect ${\rm SNR} \ge 5$
in the three smallest $k$ bins with an  SNR in excess of  $10$ in the
bin centered at $k \approx 0.07  \, {\rm Mpc}^{-1}$.  The predicted
SNR values are somewhat increased at $z_c = 2.55$ (right panel) where
we expect ${\rm SNR} \ge 5$ in the three smallest $k$-bins for both 
$B= 30$ and $60 \, {\rm MHz}$. 
     
\section{Summary and Conclusions}

We have carried out a Fisher matrix analysis to study the prospects
of measuring the cross-correlation of the redshifted HI 21-cm signal
and the Lyman-$\alpha$ forest. For the redshifted 21-cm signal we have
considered   an upcoming radio-interferometric array OWFA 
\citep{ali-bharadwaj2014}  whereas for 
the Lyman-$\alpha$ forest we have considered the currently available 
DR14 of SDSS \citep{SDSS14}. As discussed in \citep{guha-sarkar2013,guha-sarkar2015}, the 
cross-correlation signal holds advantages over the auto-correlation of
the redshifted HI 21-cm signal and the Lyman-$\alpha$ forest in that the
problem of foregrounds is less severe for the cross-correlation signal
as compared to the  21-cm auto-correlation signal. Further,  the
cross-correlation signal is expected to be less sensitive to the
discrete  QSO sampling as compared to the Lyman-$\alpha$ forest 
auto-correlation signal. Both these advantages make the
cross-correlation signal an important cosmological probe from the
observation point-of-view.

OWFA PII is expected to operate with a bandwidth of around 
$B = 30 \, {\rm MHz}$ centered at $326.5 \, {\rm MHz}$ which
corresponds to $z_c=3.35$. The  quasar number distribution from
DR14 of SDSS \citep{SDSS14} peaks at $z = 2.25$ and falls off at
higher redshifts. The number
of quasars available for the cross-correlation analysis at $z_c =
3.35$ is quite  low. In view of this, we have also explored the
possibility of increasing the number of available quasars by 
tuning the OWFA observational frequency to $z_c = 3.05$
and $2.55$ which are closer to the peak.  In addition to this, we have
also considered the possibility of increasing the bandwidth to 
$B = 60 \, {\rm MHz}$.

We find that it is not possible to measure  the amplitude of the 
cross-correlation signal with ${\rm SNR} \sim 5$  using observations
in a single OWFA field of view for any of the redshifts $z_c$ and bandwidths $B$
considered here (see Figure \ref{fig:snr_mnd}).  The SNR increases
very slowly at  $t \geq 200$ hrs and it is not 
meaningful to consider deeper observation in a single FoV. We have
considered $t =200$ hrs  observations each in $N_p=25$ independent
fields of view. We find (Table~\ref{tab:snr_mnd}) that a $6 \, \sigma$
detection is possible for $z_c=3.35$ and $B=30 \, {\rm MHz}$. The SNR
increases if $z_c$ is reduced or $B$ is doubled, and we ${\rm SNR}
\sim 17$ for $z_c=2.25$ and $B=60 \, {\rm MHz}$. 

We have considered the possibility of jointly measuring $\beta_F$ and
$\beta_T$, marginalizing over the amplitude of the cross-correlation
signal. We find that the predicted errors in these two parameters are
quite large and highly anti-correlated. Identifying combination of
these two parameters which could be independently measured, we find
that it  is at best  possible to measure 
$\beta_F^{0.83}\beta_T^{0.55}$ at an ${\rm SNR} \sim 3$ for $z_c =
2.55$ and $B = 60 \, {\rm MHz}$. In order to improve the prospects
further, we have considered the possibility of combining this with independent
measurements of $\beta_T$ from  the 21-cm auto-correlation signal
with an observing time of $1000$ hrs each in $N_p = 5$ 
independent FoVs. We have ${\rm SNR} \approx 4.1 $ and $1.67$ for
$\beta_T$ and $\beta_F$ respectively at $z_c=3.35$ with $B=30 \, {\rm
  MHz}$. In the best case the respective SNR values are $12.8$ and $5$ 
at $z_c=2.55$ with $B=60 \, {\rm   MHz}$. The parameter $\beta_T = f/ b_{HI}$ 
is sensitive to the cosmological model through the  growth function 
$f$ and the distribution of $HI$ modeled using $b_{HI}$. 
Several probes of cosmological structure 
formation like the CMBR observations and galaxy redshift 
surveys put stringent bounds on the growth function $f$.
Hence, the measurement of $\beta_T$ shall immediately 
constrain $b_{HI}$, throwing valuable insights about the 
HI distribution in the post reionization epoch.

Finally we have considered the possibility of constraining 
the shape of the cross-correlation  power spectrum by measuring its
amplitude in different $k$ bins. We find that for the $z_c$ values
considered here  the SNR peaks at $k \approx 0.07 \, {\rm 
  Mpc}^{-1}$ and $0.1 \, {\rm Mpc}^{-1}$ for $B = 30 \, {\rm MHz}$ and
$60 \,{\rm MHz}$ respectively. For $B = 30 {\rm MHz}$ at $z_c= 3.35$ a 
${\rm SNR} \ge 5$ measurement is not possible in the any of the
$k$-bins, however a ${\rm SNR} \ge 3$ measurement is possible 
in the two smallest $k$ bins. However a ${\rm SNR} \ge 5$ measurement
is possible  in the two and three smallest $k$ bins at $z_c=3.05$ and
$2.55$ respectively.  For $B=60 \, {\rm MHz}$ a ${\rm SNR} \ge 5$
measurement is possible at all three $z_c$ values, the SNR values also
increase if $z_c$ is reduced to a value closer to the peak of the QSO
distribution. 

The entire analysis here is based on the assumption that 21-cm 
foregrounds have been modeled and removed prior to the
cross-correlation, and the residual foreground contamination is
restricted to the 
$k_{\parallel}=0$ modes which have been discarded for the Fisher
matrix. While the foreground contamination is expected to be less
severe for the cross-correlation as compared to the auto-correlation,
it is quite possible that the foreground contamination will extend
beyond  the $k_{\parallel}=0$  modes causing a degradation of the
SNR. 

In conclusion we note that a $6 \, \sigma$ detection of the
cross-correlation signal will be possible with the upcoming OWFA PII
using $200$ hrs observation each in $25$ independent pointing
directions. However little else can be done beyond this using the
cross-correlation signal at $z_c=3.35$ with $B=30 \, {\rm MHz}$. 
The SNR of the cross-correlation detection
can be significantly enhanced to $\sim 17$ if the OWFA observing
redshift is tuned to $z_c=2.55$ which is closer to the peak of the QSO 
distribution and the OWFA bandwidth is increased to $60 \, {\rm
  MHz}$. It is then possible to individually measure $\beta_T$ and
$\beta_F$ with ${\rm SNR} \ge 5$ by combining the cross-correlation with
21-cm auto-correlation measurements. Further, it is also possible to
measure the amplitude of the binned cross-correlation power spectrum
in several bins at $k \le 0.3 \, {\rm Mpc}^{-1}$. 

\section{Acknowledgment}

AKS would like to thank Jayaram N. Chengalur for suggesting the
problem. AKS would also like to acknowledge the help of Debanjan
Sarkar, Siddhartha Bhattacharyya and Suman Chatterjee in writing the
manuscript. AKS makes a special mention of Tirthankar Roy
Choudhury for useful discussions which helped in a better understanding
 of the subject. 

\bibliographystyle{JHEP} 
\bibliography{ref}

\end{document}